# Phase-controllable growth of ultrathin 2D magnetic FeTe crystals

Lixing Kang[1,2,§], Chen Ye[3,§], Xiaoxu Zhao[5,§], Xieyu Zhou[4,§], Junxiong Hu[6], Qiao Li[8], Qingling Ouyang[2], Jiefu Yang[1], Dianyi Hu[1], Jieqiong Chen[1], Xun Cao[1], Yong Zhang[1], Manzhang Xu[1], Jun Di[1], Dan Tian[1], Pin Song[1], Govindan Kutty[1], Qingsheng Zeng[1], Qundong Fu[1], Ya Deng[1], Jiadong Zhou[1], Stephen J. Pennycook[5], Ariando Ariando[6], Feng Miao[8], Guo Hong[9], Yizhong Huang[1], Ken-Tye Yong[2]*, Wei Ji[4]*, Xiao Renshaw Wang[3]*, Zheng Liu[1,2,7]*

[1]School of Materials Science and Engineering, Nanyang Technological University, Singapore 639798, Singapore.

[2]CINTRA CNRS/NTU/THALES, UMI 3288, Research Techno Plaza, Singapore 637553, Singapore

[3]School of Physical and Mathematical Sciences, Nanyang Technological University, Singapore 639798, Singapore.

[4]Department of Physics and Beijing Key Laboratory of Optoelectronic Functional Materials & Micro-Nano Devices, Renmin University of China, 100872, Beijing, China.

[5]Department of Materials Science and Engineering, National University of Singapore, Singapore 117575, Singapore

[6]Department of Physics, National University of Singapore, Singapore 117551, Singapore

[7]Centre for Micro-/Nano-electronics (NOVITAS), School of Electrical and Electronic Engineering, Nanyang Technological University, Singapore 639798, Singapore

[8]National Laboratory of Solid State Microstructures, School of Physics, Collaborative Innovation Center of Advanced Microstructures, Nanjing University, Nanjing 210093, China

[9]Institute of Applied Physics and Materials Engineering, Department of Physics and Chemistry, Faculty of Science and Technology, University of Macau, Macau SAR 999078, China.

[§]These authors contributed equally: Lixing Kang, Chen Ye, Xiaoxu Zhao and Xieyu Zhou

Correspondence and requests for materials should be addressed to K.-T. Y. (ktyong@ntu.edu.sg) or to W. J. (wji@ruc.edu.cn) or to X.R.W. (email: renshaw@ntu.edu.sg) or to Z.L. (email: z.liu@ntu.edu.sg)

# Abstract

Two-dimensional (2D) magnets with intrinsic ferromagnetic/antiferromagnetic (FM/AFM) ordering are highly desirable for future spintronics devices. However, the synthesis of 2D magnetic crystals, especially the direct growth on $SiO_2$/Si substrate, is just in its infancy. Here, we report a chemical vapor deposition (CVD)-based rational growth approach for the synthesis of ultrathin FeTe crystals with controlled structural and magnetic phases. By precisely optimizing the growth temperature ($T_{growth}$), FeTe nanoplates with either layered tetragonal or non-layered hexagonal phase can be controlled with high-quality. The two controllable phases lead to square and triangular morphologies with a thickness down to 3.6 and 2.8 nm, respectively. More importantly, transport measurements reveal that tetragonal FeTe is antiferromagnetic with a Néel temperature ($T_N$) about 71.8 K, while hexagonal FeTe is ferromagnetic with a Curie temperature ($T_C$) around 220 K. Theoretical calculations indicate that the ferromagnetic order in hexagonal FeTe is originated from a concomitant lattice distortion and the spin-lattice coupling. This study represents a major step forward in the CVD growth of 2D magnetic materials on $SiO_2$/Si substrates and highlights on their potential applications in the future spintronic devices.

# Introduction

Recently, 2D magnets have drawn much attention because they provide an ideal platform for exploring magnetism down to atomic-layer thicknesses[1,2]. The discovery of long-range magnetic order in 2D van der Waals (vdW) crystals is vital to the understanding of the spin behavior in 2D limit and could enable novel spintronic applications ranging from molecular quantum devices to high-density data storage devices[3-5]. Different from conventional magnetic thin films and bulk counterparts, these intrinsic 2D magnetic crystals show novel properties and often bring about new physical phenomena. For instance, layer dependent ferromagnetism was observed in $CrI_3$ and $Cr_2Ge_2Te_6$ nanosheets at low temperatures with an out-of-plane anisotropy[3,6]. Further, ferromagnetic ordering with Curie Transition temperature ($T_c$) >300K was induced in $Fe_3GeTe_2$ extrinsically by an ionic gating method[7]. In addition, long-distance magnon transport was also detected in antiferromagnet $MnPS_3$ crystal[8].

However, there are two issues that hinder the practical applications of 2D magnets. First, some 2D magnets are extremely unstable (*e.g.* $CrI_3$). Therefore, glovebox-involved mechanical exfoliation or molecular beam epitaxial (MBE) has to be employed to prepare the 2D magnets in an oxygen/water free or ultra-high vacuum condition. In addition, these approaches are time-consuming and not industrially scalable[9,10]. Second, many 2D magnets show intricate phases (*e.g.* magnetic sulfides/selenides and alloy). Hence, it is challenging to identify an exact phase of the 2D magnets and provide a sound explanation on the origin of its magnetism. Recently, selective substrates, such as mica, was chosen for the growth of a certain phase of 2D magnets including magnetic ultrathin $Cr_2S_3$ and CrSe crystals[11,12]. Such epitaxial growth can only be applied to limited number of 2D magnets and also raises a transfer problem for the device fabrication or magnetic measurements. For instance, the transfer process inevitably introduces defects and contamination, which dramatically degrades the performance of devices. Thus, exploring new 2D magnetic compounds and developing reliable synthesis methods are urgent and necessary.

Fe-chalcogenides (FeS, FeSe and FeTe) are a new class of magnetic materials[13,14]. They show a variety of magnetic behaviors, such as ferromagnetic, ferrimagnetic and antiferromagnetic, with the changing of chalcogens[15,16]. In addition, Fe-chalcogenides also exhibit multiple structural phases with distinct properties[17,18]. For instance, the conventional phase of FeTe is an antiferromagnetic metal with a tetragonal crystal structure and $T_N \approx 70$ K[16]. Recently, density-functional theory (DFT) calculations predicted an intriguing hexagonal FeTe phase, which is weak magnetic with a large $T_C$ value and lattice distortion might account for the differences in magnetic structures[19]. Since the magnetic properties can be significantly affected by the change in structural phases, controlled synthesis of FeTe with tunable structural phases is needed. Thus far, the synthesis of 2D FeTe crystals by CVD method is rarely reported, much less the phase control[20].

In this work, we have directly synthesized antiferromagnetic and ferromagnetic FeTe on $SiO_2$/Si substrates. By controlling the growth temperature, ultrathin 2D layered tetragonal FeTe nanoplates and non-layered hexagonal FeTe nanoplates were selectively obtained as square or triangular shapes, with thickness down to 2.8 and 3.6 nm, respectively. The grain size of FeTe flakes is up to dozens of micrometers, suggesting they are high-quality single crystals. Furthermore, the antiferromagnetic behavior of tetragonal FeTe is exhibited with a $T_N$ of 71.8 K, while the ferromagnetic hexagonal FeTe with a $T_C$ of 220 K. Theoretical calculations indicate that the structural distortion is responsible for the observed ferromagnetism.

# Results

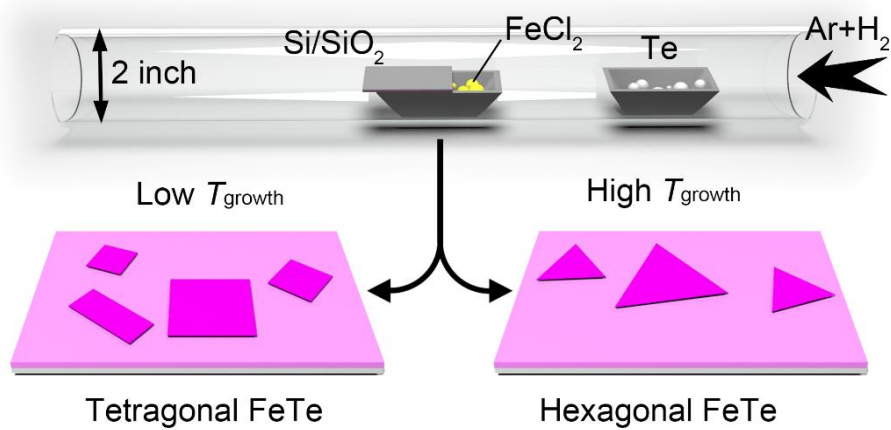

**Fig. 1** The schematic view for the iron tellurides growth process

**Phase-tunable growth of FeTe crystals.** Figure. 1 schematically illustrates the setup of the ambient pressure CVD system for the growth of FeTe crystals on $SiO_2$/Si substrates. Briefly, ferrous chloride ($FeCl_2$) and tellurium (Te) powders were used as reactants. During the CVD process, the Ar/$H_2$ mixtures carry a certain amount of tellurium vapor at a fixed heating temperature of tellurium source ($T_{Te}$) to react with evaporated $FeCl_2$ at different growth temperature ($T_{growth}$). Details about the sample synthesis are provided in the Method section. As shown in Supplementary Fig. 1, the tetragonal FeTe belongs to *P4/nmm* space group owning a layered structure in which Fe atom layer and the double slabs of Te atom layers are interlaced in the interlayer direction (Supplementary Fig. 1a). While, hexagonal FeTe has a non-layered structure which belongs to $P6_3/mmc$ space group. This structure can be regarded as the closed-packed Fe atomic planes alternatively occupy the octahedral vacancies created by the *AB*-stacked Te atomic planes. (Supplementary Fig. 1b).

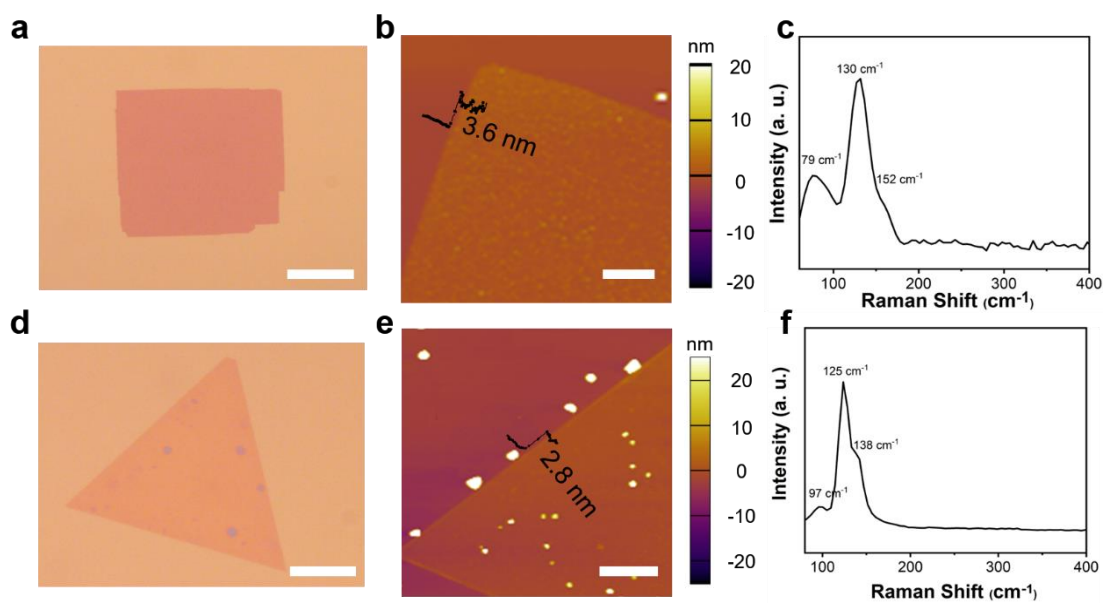

**Fig. 2** Morphological and structural characterization of the as-synthesized FeTe samples. **a, d** Typical optical images of as-grown tetragonal (**a**) and hexagonal (**d**) FeTe crystals on SiO$_2$/Si substrates. **b, e** AFM images the square FeTe nanoplate with a thickness of 3.6 nm and the trigonal FeTe nanoplate with a thickness of 2.8 nm. **c, f** Typical Raman spectrum of tetragonal and hexagonal FeTe nanoplates. Scale bars: 20 μm in **a**, **d**; 2 μm in **b**; 3 μm in **e**.

One of the most important features of FeTe is its phase tenability which originates from the formation energy difference between the hexagonal and tetragonal phases in FeTe[13,14]. Theoretical calculations have predicted that the hexagonal FeTe is the most thermodynamically favorable phase[16,19,21]. Thus, the growth temperature in the CVD process is essential to realize the phase transition. We found that the phase of FeTe was sensitive to its growing temperature ($T_{growth}$). As shown in Fig. 1, at a relative high temperature, hexagonal FeTe will be the dominant phase. Decreasing $T_{growth}$ will lead to the formation of tetragonal FeTe. A clear trend can be seen from the following optical images (Fig. 2a, d and Supplementary Fig. 2a-c). Notably, square-shaped FeTe crystals with an edge size of about 40 μm were obtained at a $T_{growth}$ of 530 ℃ (Fig. 2a). At 550 ℃, the obtained nanosheets remain square-like, with slightly increased thickness estimating by the optical contrast (Supplementary Fig. 2a). However, a mixed square and hexagonal shapes were observed at 570 ℃ (Supplementary Fig. 2b). When $T_{growth}$ was further increased to 590 ℃, the resulting FeTe nanoplates exhibited homogeneous triangular shape nanoplates with domain sizes exceeding 60 μm (Fig. 2d). Further increasing $T_{growth}$ to 610 ℃ yielded thicker nanoplates again (Supplementary Fig. 2c). The thickness, grain size and phase structure distribution histogram obtained at different temperature are shown in Supplementary Fig. 2d-f. The phase evolution of FeTe flakes from tetragonal to hexagonal highly relies on $T_{growth}$ (Supplementary Fig. 2f). These results further prove that maintaining a relative high temperature is essential for obtaining the thermodynamically stable hexagonal FeTe phase. Similarly, the reported literatures have also shown that the temperature and amount of precursor play a key role for phase transformations in 2D materials[22-25].

**Structural characterization of FeTe crystals.** Further characterizations were performed to investigate the morphology and composition of the as-obtained FeTe crystals. The atomic force microscopy (AFM) image in Fig. 2b shows an individual tetragonal FeTe flake with a thickness of 3.6 nm. As for hexagonal FeTe, the thickness can be tailored down to 2.8 nm (Fig. 2e), which are extremely thin for a non-layered material. Fig. 2c shows the Raman spectra of the as-grown tetragonal FeTe flakes. Two obvious Raman peaks were located at 130 and 152 cm$^{-1}$, corresponding to the $E_g$ and $A_{1g}$ modes of tetragonal FeTe, which is in good accordance with the previous studies[26]. A similar result with different peak position was also observed in the Raman spectra of the hexagonal FeTe crystal (Fig. 2f). Supplementary Fig. 3 shows optical images and corresponding Raman intensity maps for tetragonal and hexagonal FeTe flakes, respectively, suggesting the high crystallinity and uniformity of the FeTe crystals. In addition to Raman characterization, X-ray photoelectron spectroscopy (XPS) was used to analyze the chemical composition of the as-grown FeTe nanosheets. The XPS results (Supplementary Fig. 4) demonstrate the pure phase of FeTe without oxidation or any residual chlorine.

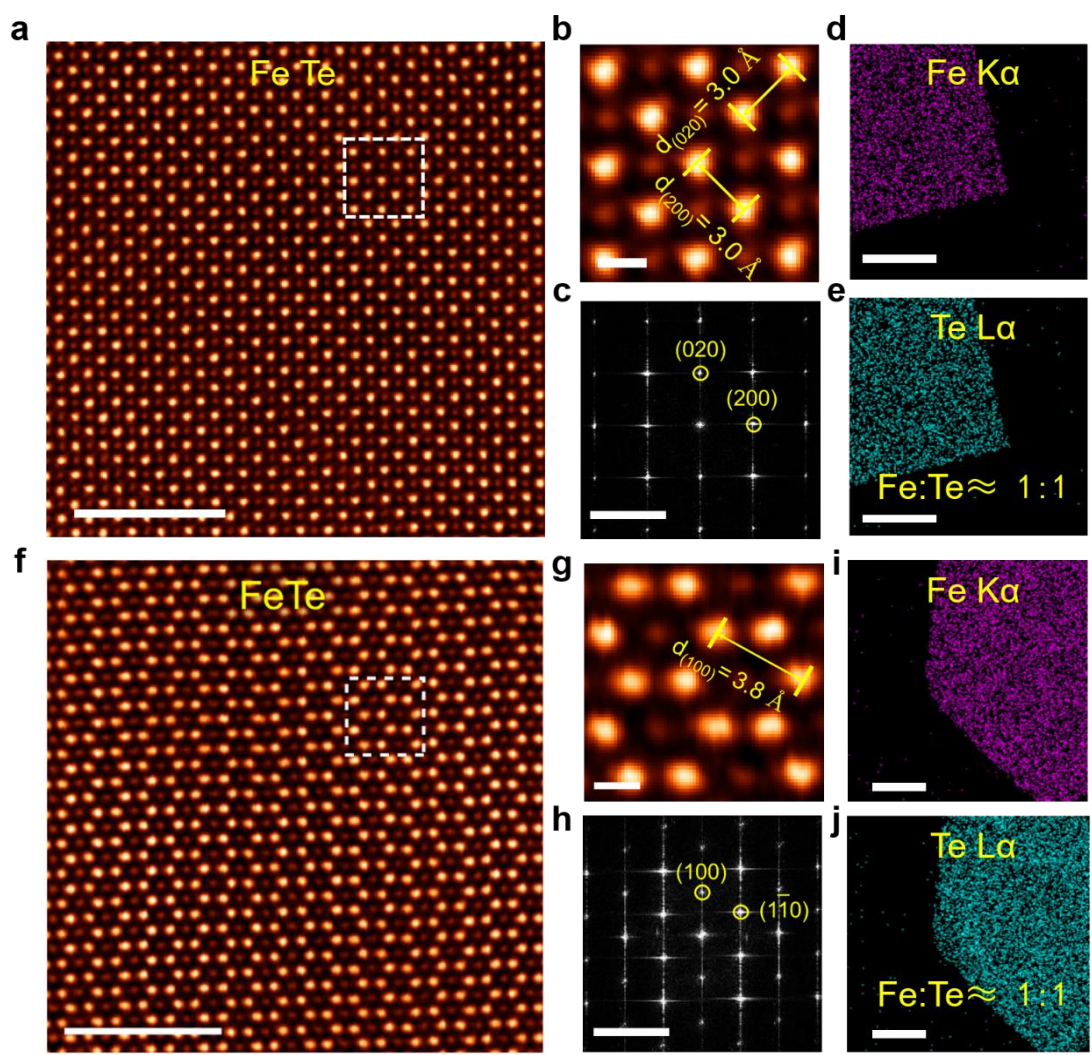

**Fig. 3** Atomic morphology of tetragonal and hexagonal shaped FeTe crystals. **a, f** Atomic-resolution STEM-ADF images of (**a**) tetragonal, and (**f**) hexagonal shaped FeTe crystals. **b, g** The magnified STEM images from the white box regions in (**a, f**), respectively. **c, h** Corresponding FFT patterns of (**a, f**), respectively. **d, e, i, j** EDS mapping of (**d, e**) tetragonal, and (**i, j**) hexagonal shaped FeTe crystals. Scale bars: 5 nm in **a, f**; 0.5 nm in **b, g**; 5 nm$^{-1}$ in **c, h**; 1 μm in **d, e** and 500 nm in **i, j**.

To probe the atomic structural differences of CVD grown tetragonal and hexagonal shaped FeTe crystals, aberration-corrected scanning transmission electron microscopy–annular dark field (STEM-ADF) imaging was applied. The image contrast in STEM-ADF image is intimately tied to the Z atomic number varying approximately as $Z^{1.6-1.7}$, thereby the Z-contrast STEM image is widely employed to identify the atomic structures in 2D materials[27,28]. The FeTe crystals were transferred from the SiO$_2$/Si substrate to TEM grids as shown in Supplementary Fig. 5, through a routine wet transfer protocol[29]. A typical atomic-resolution STEM-ADF image of the tetragonal shaped FeTe crystal along the [001] zone axis was depicted in Fig. 3a. Following the intuitive STEM image, it can be seen that in line with the macroscopically manifested tetragonal crystal. The tetragonal FeTe crystals comply with a P$_{4g}$ wallpaper group symmetry. The in-plane Te-Te and Fe-Fe bond is estimated ~3.0 Å as verified by the zoom in STEM image (Fig. 3b). No discernible extended defects are spotted in the STEM images confirming the as-grown tetragonal shaped FeTe crystals are highly crystalline. In addition, the corresponding fast Fourier transform (FFT) pattern reveals singlet set of spots which further corroborate the high crystallinity of the crystal (Fig. 3c). Apparently, it is a typical layered material where the interlayer bonding is raised by weak Te-Te interaction. The Fe and Te elemental distribution as suggested by EDS mapping (Fig. 3d,e) is homogenous throughout the entire crystal, and the chemical stoichiometry is calculated as FeTe.

In stark contrast to tetragonal shaped FeTe crystal, the STEM image (Fig. 3f) of hexagonal shaped FeTe crystal reveals an in-plane six-fold symmetry along the [001] zone axis. The in-plane Te-Te or Fe-Fe bond is calculated as 3.8 Å according to the enlarged STEM image (Fig. 3g). In parallel, no structural defects or stacking faults are observed throughout the flakes suggesting the hexagonal shaped FeTe crystal is highly crystalline. The single crystallinity is further verified by the corresponding hexagonal shaped FFT pattern (Fig. 3h) where only one set of spots can be observed. Structurally, closed-packed Te atomic planes take periodic ABAB stacking order, and the closed-packed Fe atomic planes alternatively occupy the octahedral vacancies created by the AB-stacked Te atomic planes. In the other words, the hexagonal shaped FeTe takes a periodic AcBc stacking registry where small and capital letters represent Fe and Te closed-packed atomic planes, respectively. Hence, the hexagonal shaped FeTe is no longer a layered material. The chemical composition of hexagonal shaped FeTe crystal was further analyzed by EDS mapping (Fig. 3i,j). The elemental distribution of Fe and Te elements is uniform and homogeneous. In addition, the chemical stoichiometry of Fe and Te is ~ 1:1.

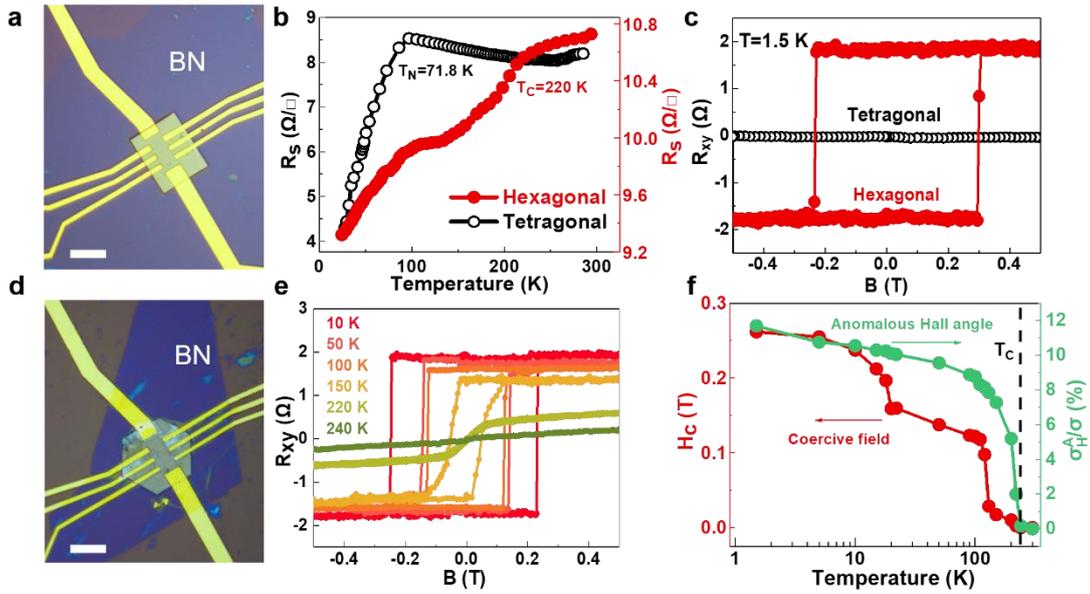

**Fig. 4** Magnetic characterization of the FeTe nanosheets. **a, d** Optical images of Hall-bar devices of individual tetragonal (**a**) and hexagonal (**d**) FeTe nanoplate protected by a thin h-BN layer. **b** The temperature dependence of longitudinal sheet resistance ($R_S$) of tetragonal and hexagonal flakes. **c** The magnetic field dependence of the Hall resistance ($R_{xy}$) at 1.5 K. **e** Temperature-dependent anomalous hall effect (AHE) of the hexagonal FeTe. **f** Temperature-dependent of the anomalous Hall angle $\sigma_H^A/\sigma$, of the hexagonal FeTe. Scale bars: 20 μm in **a, d**.

**Magnetoelectric characterization of FeTe samples on SiO$_2$/Si.** While magnetic signal can be directly detected by bulk-sensitive technique, such as vibrating sample magnetometer (VSM) and superconducting quantum interference device (SQUID), magnetoelectric measurement is more suitable to characterize the weak magnetism from low dimensional nanoflakes[2,12]. For instance, bulk sensitive techniques inevitably detect the signal from magnetic contaminations, because ultrathin samples' weak magnetic signal is often comparable or even weaker than that of contaminations obtained from various sources, such as substrate, glue, magnetic particles from laboratory environment. Fig. 4a, d shows the optical images of tetragonal and hexagonal FeTe Hall-bar devices, which were fabricated directly onto the SiO$_2$/Si substrates and well-protected by the hexagonal boron nitride (h-BN) capping layer. We should note that tellurides are susceptible to ambient degradation. Our FeTe crystals usually can only be stable existence in the air for half an hour (Supplementary Fig. 6). So, FeTe samples were always immediately moved to an Ar glovebox after the growth process. The glove box (Supplementary Fig. 7) is equipped with an optical microscope with very long-distance objectives, a spin coater, a heating stage and other required elements for the transfer, PMMA coating and h-BN encapsulation process. The details are described in the methods. We did reduce the exposure time of FeTe samples in the air as much as possible to avoid the degradation. The temperature dependence of longitudinal sheet resistance ($R_S$) of tetragonal and hexagonal flakes is depicted in Fig. 4b. A sharp decline of $R_S$ was observed at 71.8 K for the tetragonal FeTe (black curve in Fig. 4b), corresponding to a non-magnetic to AFM transition. Our temperature-dependent magnetic moment results in Supplementary Fig. 8 also confirm the AFM behavior with $T_N \sim$ 70K, below which the spontaneous magnetization of FeTe lattice exceeds over the thermal fluctuation-induced net magnetic moment. They are in accordance with the reported Néel temperature of the FeTe with a Fe:Te chemical stoichiometry of ~ 1:1[30]. In addition, we also investigated the thickness-dependent magnetism of layered tetragonal FeTe. Supplementary Fig. 9 shows the temperature-dependent resistance of three different thicknesses, 32 nm, 19 nm and 6 nm. As the thickness decline, the resistance is increased by two orders of magnitude. The magnetic phase transition is observed in 32 nm and 19 nm tetragonal FeTe flakes at $T_N$ = 67 K and 50 K, suggesting the $T_N$ would decrease as the thickness decline, which is consistent with a weak interlayer coupling[6].

The hexagonal FeTe (red curve in Fig. 4b) goes through a phase transition from paramagnetic (PM) to ferromagnetic (FM) phase at a $T_C$=220 K, below which the anomalous Hall effect (AHE) emerges. Fig. 4c shows the Hall resistance, $R_{xy}$, of both tetragonal and hexagonal FeTe devices at 1.5 K within ±0.5 T magnetic field range. The tetragonal phase exhibits a linear Hall effect with a carrier density $n_s$=4.993×10$^{16}$ cm$^{-2}$. The hexagonal FeTe exhibits an AHE with a clear hysteresis loop with a coercive field $H_c$=0.26 T and the saturated jump of $|\Delta R_{xy}|$≈2 Ω at 1.5 K. This confirms the hexagonal FeTe is ferromagnetic below $T_C$=220 K. Fig. 4e demonstrates that the temperature-dependent saturation and $H_c$ gradually decreases as the temperature increases from 1.5 to 220 K, above which the AHE vanishes.

Anomalous Hall angle, (AHA) $\theta_{AH} = \sigma_{xy}^A/\sigma_{xx}$, reflecting the efficiency of the transition from the normal current into the anomalous Hall current, is a critical parameter for quantitively analyzing the relative strength of AHE[31]. The system with a strong intrinsic AHE is more suitable for spintronic application and more likely to achieve the quantum anomalous Hall effect (QAHE)[32, 33]. Fig. 4f shows the temperature-dependent $H_c$ and $\theta_{AH}$, both gradually increase as the temperature decreases and reach the highest value of $H_c$=0.26 T and $\theta_{AH}$ =11.6% at 1.5 K. However, instead of developing smoothly, we observed that the coercive field were enhanced under 110 K. We suspect that is ascribed to the faster process of spin flip, presenting as the jump of $R_{xy}$ saturation transits from gradual to steep when T<110 K. This requires further study to confirm the origin of this unexpected enhancement.

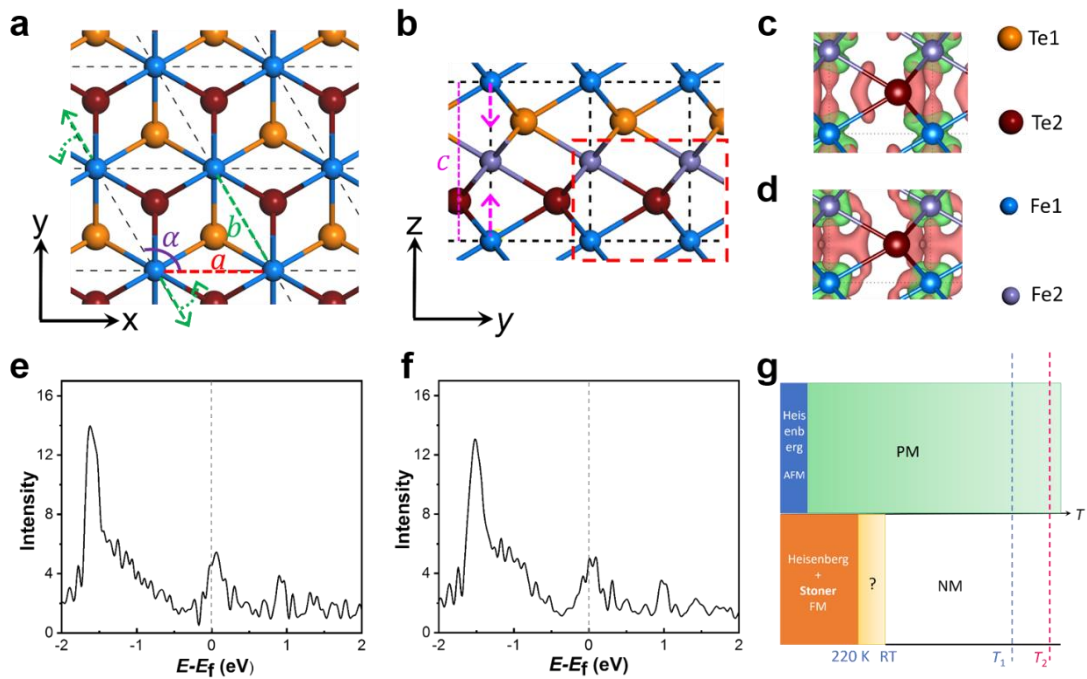

**Fig 5. Structure and magnetism of undistorted and distorted hexagonal FeTe. a** Top view of hexagonal FeTe. Lattice constants *a* and *b* are marked with the red and green dashed lines, respectively, while the purple arc represents angle *α* between the two lattice vectors. Green dashed arrows show the directions of lattice reshaping, *e.g.* extended *b* and enlarged *α*, in the distorted lattice. **b** Side view of hexagonal FeTe. Lattice constant *c* is marked with the pink dashed line while the two pink dashed arrows denote a shrunk *c* in the distorted structure. **c** and **d** Atomic differential charge densities of the undistorted and distorted structures in their most stable magnetic orders with an isosurface of 0.006 $e$/Bohr$^3$. The plotting region is marked in the red dashed rectangle in **b**. Here, red and green isosurface contours correspond to charge accumulation and reduction after Fe and Te atoms bonding together, respectively. **e** and **f** total density of states of *d* orbitals for undistorted and distorted hexagonal FeTe, respectively. **g** Phase diagram of magnetism of the tetragonal and hexagonal phases as a function of temperature. The tetragonal (upper) and hexagonal (lower) phases are separated by the lateral axis of temperature. Blue and red dashed line represent $T_{growth}$ of the tetragonal ($T_1$) and hexagonal ($T_2$) phases, respectively. Here, the paramagnetic phase is modeled by a Neel AFM configuration for the tetragonal phase above $T_c$, while it is, most likely, a NM one for the hexagonal phase.

**Insight of the origin of magnetic order in FeTe crystals**. Density functional theory calculations were carried out to determine the magnetic ground state of the hexagonal FeTe. Schematic top and side views of the hexagonal FeTe structure are shown in Fig. 5a,b. Our non-magnetic calculation reveals lattice constants *a*=3.90 Å and *c*=5.36 Å, comparable with the STEM (Fig. 3g) and electron diffraction (Supplementary Fig. 10) values of 3.8 Å and 5.2 Å measured at room temperature, respectively. If Fe cations are allowed to have local magnetic moments, however, lattice constant *a* elongates to roughly 4.2 Å regardless its long-range magnetic ordering. Together with the $T_C$ of ~ 220 K (Fig. 4b), these results suggest the structure observed in STEM is, most likely, a non-magnetic but not a para-magnetic hexagonal FeTe. In terms of the tetragonal phase, however, the measured value of *a*=3.0 Å is closer to our theoretical value of Néel AFM phase of *a*=2.90 Å, but much larger than the non-magnetic value of *a*=2.71 Å. A Néel AFM configuration can mostly simulate properties of the corresponding paramagnetic phase, which implies local magnetic moments of the tetragonal phase persist at room temperature, consistent with previous calculations and experiment measurements[16]. In light of this, we can infer that the tetragonal and hexagonal phases are, most likely, paramagnetic (PM) and non-magnetic (NM), respectively at room temperature and during the synthesis process (Supplementary Fig. 11).

We first consider a local magnetic moment picture for the ferromagnetism of the tetragonal phase. Significant magnetostriction was observed in the FM-FM (both intra- and inter-layer FM, *a*=4.23 Å, *c*=5.81 Å) and FM-AFM (intralayer FM and interlayer AFM, *a*=4.19 Å, *c*=5.83 Å) configurations, which are yet to be confirmed by low temperature STEM experiments. Details of all considered magnetic orders are available in Supplementary Fig. 12. Among these configurations, the FM-AFM order (Fig. 5c) is, however, over 100 meV/Fe more favored than other orders for a perfect hexagonal FeTe (Supplementary Table 1). A structural distortion further lowers the energies of all magnetic orders. As a result of the distortion, lattice constant *c* shows a considerable shrink from roughly 5.8 Å to 5.45 Å (Fig. 5b), associated with a stretchered lattice constant *b* (from 4.19 Å to 4.29 Å) and an increased angle *α* (from 120.0° to 121.4°) (Fig. 5a). The distorted structure leads to a shorter interlayer Fe-Fe distance and thus enhances their FM direct exchange, which favors the FM-FM order in a slightly distorted structure (Fig. 5d and

Supplementary Table 1). These results indicate the structural distortion might play a paramount role in the observed ferromagnetism.

Spin-exchange parameters for both the undistorted and distorted structures are listed in Supplementary Supplementary Fig. 12 and Supplementary Table 2, which predict a $T_c$ value around 290 K using Metropolis Monte Carlo (MC) simulation in a 3-layer 3D Ising lattice. Given the fact that MC simulation of the Ising model usually two or three times overestimates[34] $T_c$, the hexagonal phase is expected to have a $T_c$ value around 100 ~ 150 K in consideration of the Heisenberg local exchange picture, substantially lower than the measured value of ~220K. This low $T_c$ value indicates another mechanism may corporately results in such a high $T_c$ value of 220 K. Stoner ferromagnetism mechanism is another likely reason for the FM. Non-magnetic calculations were performed to calculate the DOSs using the FM geometries. Fig. 5e,f show the plots of the total $d$-orbital DOSs of the undistorted and distorted structures, respectively, both of which meet the Stoner criterion, *i.e.* $(U'/N)\rho(E_{df})>1$, compellingly indicating the Stoner ferromagnetism in this metallic system. In additional, the distorted structure shows a larger $(U'/N)\rho(E_{df})$ value (1.95) than the undistorted structure (1.55), evidencing a stronger Stoner ferromagnetism in distorted FeTe (see Supplementary Fig. 13 for details). The Stoner FM could also explain the coexistence of the high $T_c$ of 220 K and the non-magnetic state observed at RT. In summary of these result, we conclude that the hexagonal phase exhibits a Heisenberg plus Stoner FM below 220 K, NM above RT and, most likely, NM between 220 K and RT, as illustrated in Fig. 5g.

Vibrational frequency calculations were performed to compare free energies of different phases. The free energy plot (Supplementary Fig. 14) shows, however, no thermally driving phase transition occurred from 0 K to 1000 K for the bulk phases. In light of this, we believe the controlling of phases is a result of either kinetic reasons or thermally driving transition at the thin-film limit. Our results of surface energy calculation (Supplementary Fig.15) show an ultra-low surface energy (1.9 meV/Å$^2$) of the Te-terminated hexagonal slab in a Te-rich condition, roughly an order of magnitude smaller than those of other slabs. Thus, the Te-terminated hexagonal slab is the most energetically favored one among all hexagonal and tetragonal slabs in the Te-rich growth condition that could be reached by increasing the Te source temperature, as we demonstrated in our experiments. In addition, our vibrational frequency calculations indicate that Te-terminated hexagonal FeTe mono- and few-layers are, exceptionally, unstable and tend to transform into Fe-Te clusters at 0 K, with imaginary frequencies of *e.g.* 79.21 cm$^{-1}$ and 188.83 cm$^{-1}$ for the monolayer. Given the experimental synthesized hexagonal phase, we infer that the hexagonal few-layers is, most likely, stabilized by non-harmonic effects, which suppress the phonon softening and are more pronounced at high temperatures[35]. This partially explains the reason why the hexagonal phase is a high temperature phase.

## Discussion

In conclusion, a facile and effective strategy for growing ultrathin 2D magnetic FeTe crystals with tunable structural phases was developed. We can selectively obtain layered tetragonal FeTe nanoplates and non-layered hexagonal FeTe nanoplates as square or trigonal geometries, with thickness down to 2.8 and 3.6 nm, respectively. Systematic RM, AFM and STEM studies revealed fine structural information of these FeTe flakes and confirmed they were single crystals with high crystallinity. Furthermore, the magnetism based transport measurements showed tetragonal FeTe crystal was antiferromagnetic with $T_N \approx 71.8$ K. While the hexagonal FeTe flakes exhibited obvious ferromagnetic properties with a transition temperature close to 220 K. Detailed calculated results indicated a structural distortion in hexagonal FeTe is responsible for the observed ferromagnetism. Our work paves the way for controlled growth of other 2D magnetic materials by the CVD method and provides a tunable material system for investigating 2D magnetism and potential applications in future spintronic devices.

# Methods

**Atomically thin FeTe crystals synthesis.** 2D polymorphous FeTe flakes were synthesized on $SiO_2$/Si substrates by an ambient pressure CVD method. The reaction process was conducted in a 1.2-meter length, 2-inch outer diameter quartz tube heated by a three-zone furnace (Thermcraft (XST-3-0-18-2V2)) which had three independently controlled temperature zones. Te powder (99.997%, Sigma Aldrich) and ferrous chloride ($FeCl_2$, 99.9%, Sigma Aldrich) were used as solid precursors and reactants. The Te powder (in a quartz boat) was placed near to the gas inlet in the first heating zone while the ferrous chloride powder, loaded into the third heating zones, was placed in an alumina boat covered with an inclined downward facing $SiO_2$/Si substrate. The second heating zone was kept the same condition as the first heating zone. Prior to the growth process, the furnace was purged by the flowing of 300sccm high-purity Ar gas for 5 min. Then, 100 sccm Ar and 10 sccm $H_2$ mixtures were introduced into the CVD system. The tellurium was heated to 520 ℃ in 20 min while the growth zone of ferrous chloride was gradually increased to the different temperature (520~630 ℃) in 20 min. Set temperature for all the heating zones would be kept for another 10 min for growth temperature. After synthesis, the furnace was rapidly cooled to room temperature with the assistance of electric fans. Obvious etching phenomena were observed when the furnace was natural cooling without any external perturbations for both tetragonal and hexagonal FeTe flakes (Supplementary Fig. 16).

**Preparation of STEM Sample**: A thin PMMA film was spin-coated on top of FeTe layers grown on 285 nm $SiO_2$/Si and cured at 120 ℃ for 2 min in an Ar glove box. The PMMA/FeTe film was detached from the substrate by etching $SiO_2$ layer in 1 mol/L KOH solution. After rinsed in DI water, it was scooped up by a TEM grid and dried in an Ar glove box. Finally, PMMA was removed by acetone. The whole transfer process was performed in an Ar glove box due to the ultrathin tellurides were sensitive to oxidation.

**Devices fabrication and transport measurement.** The tetragonal and hexagonal FeTe flakes grown on $SiO_2$/Si substrate were quickly moved to an Ar glove box and spin coated with 950K PMMA (MicroChem) at 3000 r.p.m. for 1 min. After leaving the PMMA layer consolidating on a 120 ℃ hotplate for 5 min in the Ar glove box, small markers were fabricated using standard e-beam lithography (Nova nanoSEM 230 with digital pattern generator Nabity-nanometer pattern generation system (NPGS)) near the identified sample for subsequent fabrication of Hall-bar devices. The Ti/Au (5/70 nm) electrodes were deposited using an electron-beam evaporator (Kurt J. Lesker Nano 36 Thermal Evaporator) followed by lift-off in acetone. The electrical contacts were made by wire bonding Al wires on a Hall-bar structure to form a standard six-terminal contact pad. Transport experiments were performed in an Oxford cryostat with the temperature ranging from 300 to 2 K and the magnetic field within ±8 T. The sample resistance and Hall effect were measured by Keithley 6221 triggered with Keithley 2182 with the 21 Hz frequency in the standard four-terminal configuration. The sheet carrier concentration was determined by the Hall coefficient as $n_{2D} = 1/eR_H$, where $R_H$ is the Hall coefficient and $e$ is the elementary charge.

**Characterizations of 2D FeTe crystals.** The as-obtained FeTe nanoplatelets were further characterized by optical microscopy (Olympus BX53M), RM (WITEC alpha 300R Confocal Raman system using a 532 nm laser as the excitation source), AFM (Asylum Research Cypher Scanning Probe Microscope system with a tapping mode), XPS (Kratos AXIS Supra spectrometer with a monochromatic Al K-alpha source). ADF-STEM imaging was conducted on an aberration-corrected JEOL ARM-200F equipped with a cold field emission gun, operating at 80 kV, and an Advanced STEM Corrector (ASCOR) probe corrector.

**Details of DFT calculation.** Our density functional theory calculations were performed using the generalized gradient approximation and the projector augmented wave method[36,37] as implemented in the Vienna ab-initio simulation package (VASP) [38,39]. The Perdew-Burke-Ernzerhof (PBE) [40] functional with the density dependent dispersion correction (dDsC) [41,42] was adopted for all calculations. A uniform Monkhorst-Pack $k$-mesh of 21×21×13 was adopted for sampling over the Brillouin zone of hexagonal FeTe and a 20×20×8 mesh for the tetragonal phase. A plane-wave kinetic energy cutoff of 700 eV was used for structural relaxation while 400 eV for energy comparison. All atoms were allowed to relax until the residual force per atom was less than 0.01 eV/Å. We also consider spin-orbit coupling (SOC) in for energy comparisons among different magnetic configurations. A 2×2√3×1 orthorhombic supercell, with a $k$-mesh of 4×4×6, was used to model different magnetic orders in FeTe. An On-site Coulomb interaction energies to Fe $d$ orbitals were self-consistently calculated using a linear response method[43], reveals $U$ = 3.3 eV and $J$ = 0.5 eV.

**Derivation of spin-exchange parameters.** A Heisenberg model was used to model the magnetism of the bulk, $H = H_0 - \sum_{k=1}^{5} J_k \sum_{i,j} \vec{S_i} \cdot \vec{S_j}$, which includes two in-plane spin-exchange interactions, $J_1$ and $J_2$ and three interlayer ones, $J_3$ to $J_5$, respectively as displayed in Supplementary Fig. 12. Metropolis Monte Carlo simulation was used to estimate the $T_c$ value in a three-layer lattice using a 3D Ising model, $H = -\sum_{k=1}^{5} J_k \sum_{i,j} S_i \cdot S_j$, in which $S$=2. A 20×20×3 lattice was used with periodic boundary condition.

**Estimation of Stoner criterion.** Structural relaxation of non-magnetic configuration does not show structural distortion. We thus kept the in-plane constants $a$=3.90 Å and $b$=3.82 Å to calculate the DOS of the distorted structure. We speculate that no structural distortion are acquired at room temperature, since both geometric relaxation and STEM show no significant distortion. What we show in (b) only tend to discuss how distortion affects the DOS. Stoner parameter is defined as $ST=(U'/N)\rho(E_{fd})$. Here, $U' = \langle \epsilon_k \rangle / m_d$, $N$ is the number of unit magnetism cells which equals to 2 in this situation and $\rho(E_{fd})$ is the total density of states of all $d$ orbitals at the Fermi level for hexagonal FeTe, *i.e.* 4.54 and 4.93 states/eV/atom for undistorted and distorted structures). $\langle \epsilon_k \rangle$ is the average value of the splitting between the corresponding spin-up and spin-down bands at six high symmetry points $\Gamma$MKZAR in the BZ, which could be derived from the band structure plotted in Supplementary Fig. 13. $m_d$ is the average value of the magnetic moments on Fe atoms, i.e., 3.4 and 2.9 $\mu_B$ for undistorted and distorted FeTe.

**Acknowledgments**

This work was supported by the Singapore National Research Foundation (NRF) under NRF award number NRF-RF2013-08, Tier 2 MOE2015-T2-2-007, MOE2015-T2-2-043, MOE2017-T2-2-136, Tier 3 MOE2018-T3-1-002, AcRF Tier 2 MOE2017-T2-2-002, NRF2017-NRF-ANR002 2DPS and A*Star QTE programme. X.R.W. acknowledges supports from the Nanyang Assistant Professorship grant from Nanyang Technological University and Academic Research Fund Tier 1 (Grant No. RG108/17 and No. RG177/18) from Singapore Ministry of Education. A.A. acknowledges the support from the NUS Academic Research Fund (AcRF Tier 1 Grants No. R-144-000-391-114 and No. R-144-000-403-114) and the NRF under the Competitive Research Programs (CRP Grant No. NRF-CRP15-2015-01). G.H. acknowledges the fund of University of Macau (SRG2016-00092-IAPME, MYRG2018-00079-IAPME, MYRG2019-00115-IAPME), and the Science and Technology Development Fund, Macau SAR (FDCT081/2017/A2, FDCT0059/2018/A2, FDCT009/2017/AMJ). W.J. and D.T. also gratefully acknowledge financial support from the National Natural Science Foundation of China (Grant No. 11622437, No. 61674171 and No. 21601086), the Strategic Priority Research Program of Chinese Academy of Sciences (Grant No. XDB30000000), the Fundamental Research Funds for the Central Universities, China, the Research Funds of Renmin University of China (Grants No. 16XNLQ01 (W.J.) and No. 19XNH065 (X.Z.)), the Natural Science Foundation of Jiangsu Province (Grant No. BK20160994). Calculations were performed at the Physics Lab of High-Performance Computing of Renmin University of China, Shanghai Supercomputer Center.

**Author contributions**

Z.L., K.T.Y., X.R.W. and W.J. conceived and supervised the project. L.K. and Z.L. designed the experiments. L.K. synthesized and characterized the sample. J.H. and A.A. fabricated the devices. C.Y. and X.R.W. carried out the transport measurements and analyzed the results. X.Z. and S.J.P. did the STEM measurement and data analysis. X.Z. and W.J. performed the first-principle calculations. All the authors discussed the results and commented on the manuscript.

**Additional information**

**Supplementary information** accompanies this paper at http://www.nature.com/naturecommunications.

**Competing financial interests:** The authors declare no competing financial interests.

**Reprints and permission** information are available online at http://npg.nature.com/reprintsandpermissions/.

**Publisher's note:** Springer Nature remains neutral with regard to jurisdictional claims in published maps and institutional affiliations.